\documentclass{elsart}
\usepackage{amssymb}
\usepackage{graphicx}

\begin{document}

\begin{frontmatter}

\title{Enhancement of $B(E2)\hspace{-1.5mm}\uparrow$ and
       low excitation of the second 0$^+$ state
       near $N=40$ in Ge isotopes}

\author[a1]{M. Hasegawa},
\author[a2]{T. Mizusaki},
\author[a3]{K. Kaneko},
\author[a4,a5,a6]{Y. Sun}

\address[a1]{Laboratory of Physics, Fukuoka Dental College,
 Fukuoka 814-0193, Japan}
\address[a2]{Institute of Natural Sciences,
 Senshu University, Tokyo 101-8425, Japan }
\address[a3]{Department of Physics, Kyushu Sangyo University,
 Fukuoka 813-8503, Japan}
\address[a4]{Department of Physics and Joint Institute
 for Nuclear Astrophysics,
 University of Notre Dame, Notre Dame, IN 46556, USA}
\address[a5]{Department of Physics, Tsinghua University, Beijing 100084,
 P. R. China}
\address[a6]{Department of Physics, Xuzhou Normal University,
 Xuzhou, Jiangsu 221009, P. R. China}

\begin{abstract}
The long-standing problem of dramatic structure change near $N=40$
in Ge isotopes is investigated by means of large-scale shell model
calculation. The analysis of simulated calculations suggests a
possible understanding of the problem in terms of rapid increase
in the $g_{9/2}$ proton and neutron occupation. The observed
variation in excitation of the second $0^+$ state in
$^{70,72,74}$Ge appears to correlate closely with the $g_{9/2}$
occupations induced by strong proton-neutron interactions. The
enhancement of the $g_{9/2}$ occupancies is probably due to
correlations in the $1g2d3s$ shell.

\end{abstract}

\end{frontmatter}

\section{Introduction}\label{sec1}

Experimental information on nuclei far from stability constitutes
a challenging test for the applicability of nuclear models and
serves as a guidance for improvement of the models. Radioactive
ion beam facility plays a vital role in gathering such
information. In a recent experiment at the Oak Ridge HRIBF,
Padilla-Rodal {\it et al.} measured $B(E2;0_1^+ \rightarrow
2_1^+)$ in $^{78,80,82}$Ge \cite{Padilla}, which provides a
complete view of $B(E2)\hspace{-1.5mm}\uparrow$ data extended to
the neutron-rich region, reaching the major shell closure at
$N=50$. It was also demonstrated \cite{Padilla} that the observed
$B(E2)$ values in $^{78,80,82}$Ge can be reproduced by using a
shell model with the model space
$(p_{3/2},f_{5/2},p_{1/2},g_{9/2})$ for both protons and neutrons.
Hence this work showed that the structure change in the exotic
mass region could be microscopically studied by means of
large-scale shell model calculation, which has become available
only recently. Large-scale shell model calculations have been
applied to the study of $N=40$ magicity in Refs.
\cite{Caurier,Sorlin,Langanke1,Langanke2}.

The work of Padilla-Rodal {\it et al.} has stimulated our interest
in the long-standing problem of structure change along the Ge
isotopic chain, particularly in the vicinity of $N=40$. The
experimental $B(E2)\hspace{-1.5mm}\uparrow$ data shown in Fig. 4
of Ref. \cite{Padilla} suggest that $^{68}$Ni can be regarded as a
quasi-doubly-closed-shell nucleus, but the nature of the $N=40$
subshell closure is gradually destroyed as the proton number
outside the $f_{7/2}$ orbit increases. The variation of
$B(E2)\hspace{-1.5mm}\uparrow$ along the $N$ axis reminds us of
the old problem, much discussed in the 1970's and 1980's, of the
structure change from the $N=38$ to $N=42$ Ge isotope. References
were listed, for instance, in Ref. \cite{Fortune} published in
1987. A notable irregularity at $N=40$ was observed in the $(p,t)$
and $(t,p)$ reaction cross sections
\cite{Ardouin,Lebrun,Mordechai}, which is characterized by a
drastic drop of the second $0^+$ excitation energy in $^{72}$Ge.
This is an unusually low excited $0^+$ state, showing a similar
character as that of typical doubly-closed-shell nuclei. At
$N=40$, the drop of $E_x(0_2^+)$ appears in coincidence with a
remarkable enhancement of $B(E2)\hspace{-1.5mm}\uparrow$. These
experimental data suggest that the structure of $^{72}$Ge with
$N=40$ is very different from that of $^{68}$Ni studied in Refs.
\cite{Caurier,Sorlin,Langanke1}.

The structure changes in Ge isotopes were discussed in terms of
pairing correlation and shape change (or shape coexistence) by
using various collective models
\cite{Vergnes,Lecomte,Carchidi,Sugawara}. The interacting boson
model \cite{Duval} reproduced the observed energy levels and
$B(E2)$ values for $^{68-76}$Ge. However, a satisfactory
microscopic explanation has not been seen. Most of the early
studies in the literature limited their discussion on the
configuration change within the $fp$ shell. Contributions from the
$g_{9/2}$ orbit were very little mentioned. These early studies
are insufficient according to the new knowledge on the $N=40$
magicity obtained in Refs.
\cite{Caurier,Sorlin,Langanke1,Langanke2}, which has emphasized
that nucleons in the $g_{9/2}$ orbit play a crucial role in
determining the structure of the $N=40$ isotones. Thus, as a
necessary requirement, a microscopic study for the structure in Ge
isotopes should include the $g_{9/2}$ orbit.

\section{What problem exists?}\label{sec2}

In the previous publications \cite{Hase1,Kaneko1,Hase2}, we have
demonstrated the feasibility of a shell model description for
$^{64-68}$Ge (including odd-mass isotopes), and for some other
neighboring nuclei of the mass region. In these studies,
large-scale shell model calculations (up to dimension $2 \times
10^8$) were performed in the model space
$(p_{3/2},f_{5/2},p_{1/2},g_{9/2})$, the same model space as that
of Ref. \cite{Padilla}. The aim of the present article is twofold:
to fill the gap in the Ge isotopic chain that the previous shell
model calculations have neglected, and to look for possible
structure reasons that may explain the unsolved puzzle in the
middle of the chain around $N=40$.

The extended pairing plus quadrupole ($P+QQ$) Hamiltonian
\cite{Hase1} is employed in the present work. This Hamiltonian has
recently been proposed and tested through several shell model
applications. It should be stressed that this isospin-invariant
Hamiltonian includes strong proton-neutron interactions in both
$T=0$ and $T=1$ channel. The single-particle energies in our model
are $\varepsilon_{p3/2}$=0.00, $\varepsilon_{f5/2}$=0.77,
$\varepsilon_{p1/2}$=1.11, and $\varepsilon_{g9/2}$=2.50 (in MeV).
In the present work that deals with a long isotopic chain, we use
the $A$-dependent interaction strengths: $g_0=0.27(64/A)$ MeV for
the monopole pairing force, $\chi_2=0.25(64/A)^{5/3}$ MeV for the
quadrupole-quadrupole ($QQ$) force, and $\chi_3=0.05(64/A)^2$ MeV
for the octupole-octupole ($OO$) force. These are the same
parameters as those in Ref. \cite{Hase1}. The Hamiltonian contains
also five monopole correction terms (see Ref. \cite{Hase1}). The
effective charges are $e_{eff}^\pi=1.5e$ and $e_{eff}^\nu=0.5e$.

\begin{figure}
\begin{center}
\includegraphics[width=7.0cm,height=7.0cm]{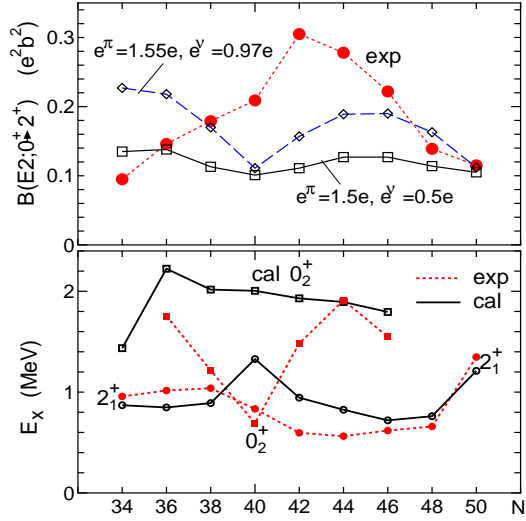}
\caption{Calculated and experimental values of
$B(E2)\hspace{-1.5mm}\uparrow$, $E_x(2_1^+)$, and $E_x(0_2^+)$ for
the Ge isotopes from $N=34$ to 50. Calculated
$B(E2)\hspace{-1.5mm}\uparrow$ with two sets of effective charge
are shown for comparison.} \label{fig1}
\end{center}
\end{figure}


Our results are presented in Fig. \ref{fig1}, where calculated
$B(E2)\hspace{-1.5mm}\uparrow$, $E_x(2_1^+)$, and $E_x(0_2^+)$ for
the Ge isotopes from $^{66}$Ge ($N=34$) to $^{82}$Ge ($N=50$) are
compared with data. The calculation reasonably reproduces
$B(E2)\hspace{-1.5mm}\uparrow$ and $E_x(0_2^+)$ at both ends of
the isotopic chain with $N$ = 34, 36 and $N$ = 48, 50, but fails
to describe those in between. Especially, the calculation could
not reproduce the increasing trend of
$B(E2)\hspace{-1.5mm}\uparrow$ over $N=40$ and the drop of
$E_x(0_2^+)$ around this neutron number. We can get an improved
agreement with the $B(E2)\hspace{-1.5mm}\uparrow$ data for
$^{70}$Ge and $^{78}$Ge (see Fig. \ref{fig1}) if we enlarge the
effective charges to $e_{eff}^\pi=1.55e$ and $e_{eff}^\nu=0.97e$
($e_{eff}^\nu=0.97e$ is taken from Ref. \cite{Padilla} and
$e_{eff}^\pi=1.55e$ is fixed so as to reproduce
$B(E2)\hspace{-1.5mm}\uparrow$ in $^{82}$Ge). However, the use of
larger effective charges destroys the good agreement for $^{66}$Ge
and $^{68}$Ge already obtained in Ref. \cite{Hase1}. Moreover, the
rise in effective charges does not seem to have any positive
effect on enhancing $B(E2)\hspace{-1.5mm}\uparrow$ at $N=40$. It
is thus clear that the problem must stem from deeper structure
reasons, which cannot be resolved simply by a global fit through
effective charges.

The above results indicate an inadequacy either in the interaction
employed in the calculation, or in the model space for
$^{70-76}$Ge in question. At present, there is no available
effective interaction better than the extended $P+QQ$ interaction
for Ge isotopes in the model space
$(p_{3/2},f_{5/2},p_{1/2},g_{9/2})$. We thus proceed our study
with the extended $P+QQ$ interaction and look for any possible
reasons for this discrepancy.

The enhancement of $B(E2)\hspace{-1.5mm}\uparrow$ in the $N=40$
isotones from $^{68}$Ni to $^{78}$Sr has been qualitatively
explained by Langanke {\it et al.} \cite{Langanke2} using the
Shell Model Monte Carlo (SMMC) approach, where a $P+QQ$
Hamiltonian (with no monopole terms) is used but their model space
is larger, which includes also the $1g2d3s$ shell. It must be
noted that the SMMC calculation gives only total $B(E2)$ strength
to excited $2^+$ states and does not give the energy of the
$0_2^+$ state. The SMMC calculation predicted a large neutron
occupation number in the $g_{9/2}$ orbit ($\langle n_{g9/2}^\nu
\rangle > 2$), in contrast to our small value
 ($\langle n_{g9/2}^\nu \rangle \sim 0.23$) obtained for
$^{72}$Ge. This gives us a hint that inclusion of the
$1g_{7/2}2d3s$ orbits may significantly enhance the role of the
$g_{9/2}$ orbit through the pairing and quadrupole correlations
when the Fermi level moves toward the $1g2d3s$ shell. The
deficiency in our calculation may be due to the insufficient model
space for the nuclei where neutrons start to occupy the $p_{1/2}$
and $g_{9/2}$ orbits.

The $g_{9/2}$ occupancy of neutrons is investigated in $^{68}$Ni
with the ordinary shell model \cite{Caurier,Sorlin}, in which the
$g_{9/2}$ orbit is excluded from the model space for protons. The
value $\langle n_{g9/2}^\nu \rangle \sim 1.2$ obtained in Ref.
\cite{Sorlin} is considerably different from the SMMC result
$\langle n_{g9/2}^\nu \rangle \sim 2.2$ \cite{Langanke2}. In
another recent work \cite{Kaneko2}, we have investigated $^{68}$Ni
in the neutron model space $(p_{3/2},f_{5/2},p_{1/2},g_{9/2})$
assuming a $^{56}$Ni core. The shell model for neutrons also gives
the value $\langle n_{g9/2}^\nu \rangle \sim 1.2$, similar to that
in Ref. \cite{Sorlin}. These shell model calculations show that
the ground state of this nucleus remains a transitional character
to a superfluid phase. It is thus interesting to investigate
whether the delicate structure changes from the $N=38$ to $N=42$
Ge isotopes are caused by strong pairing correlations.

\section{Renormalization approach within the truncated space}
\label{sec3}

We continue our study within the $2p1f_{5/2}1g_{9/2}$ space as
this model space is at present the largest possible for a full
shell model diagonalization. Within this truncated space,
contributions from the $2d1g_{7/2}3s$ orbits must be expressed in
terms of effective interaction. We have made testing calculations
with different monopole corrections, which effectively change the
neutron $g_{9/2}$ occupation. However, moderate modifications for
the monopole corrections cannot produce a significant structure
change near $N = 40$. On the other hand, it is expected that the
effects of the $2d1g_{7/2}3s$ orbits tend to strengthen the
pairing and quadrupole correlations. We therefore consider to
enlarge the quadrupole matrix element $Q(g_{9/2}g_{9/2})$ to mimic
any effect from the missing $2d1g_{7/2}3s$ orbits. Note that the
quadrupole matrix elements between the $1g_{9/2}$ and
$2d_{5/2}1g_{7/2}$ orbits are nonzero and those between the $fp$
and $gds$ shells are zero (the $2d_{5/2}$ orbit nearest to
$1g_{9/2}$ has the largest quadrupole matrix element with
$1g_{9/2}$).

Based on the above considerations, we write the effective
quadrupole matrix element for the $g_{9/2}$ orbit by multiplying a
factor $\eta$
\begin{equation}
  Q_{eff}(g_{9/2}g_{9/2}) = \eta Q(g_{9/2}g_{9/2}).
\end{equation}
Physically, an $\eta > 1$ factor strengthens the $QQ$ force with
respect to the $g_{9/2}$ orbit, which is consistent with the
expectation that the Ge isotopes become deformed at the beginning
of the $g_{9/2}$ orbit and that a strong $QQ$ force contributes to
lowering of the $g_{9/2}$ orbit. A stronger $Q(g_{9/2}g_{9/2})$
increases also the effective charges selectively for the $g_{9/2}$
nucleons. In our shell model treatment, we transform the $QQ$
force into multipole-pairing-type interactions in the
particle-particle channel. Thus effectively, a stronger $QQ$ force
with respect to the $g_{9/2}$ orbit enhances also the pairing
correlations, which allows more nucleons to jump to the $g_{9/2}$
orbit.

\begin{figure}[b]
\includegraphics[width=7.0cm,height=6.4cm]{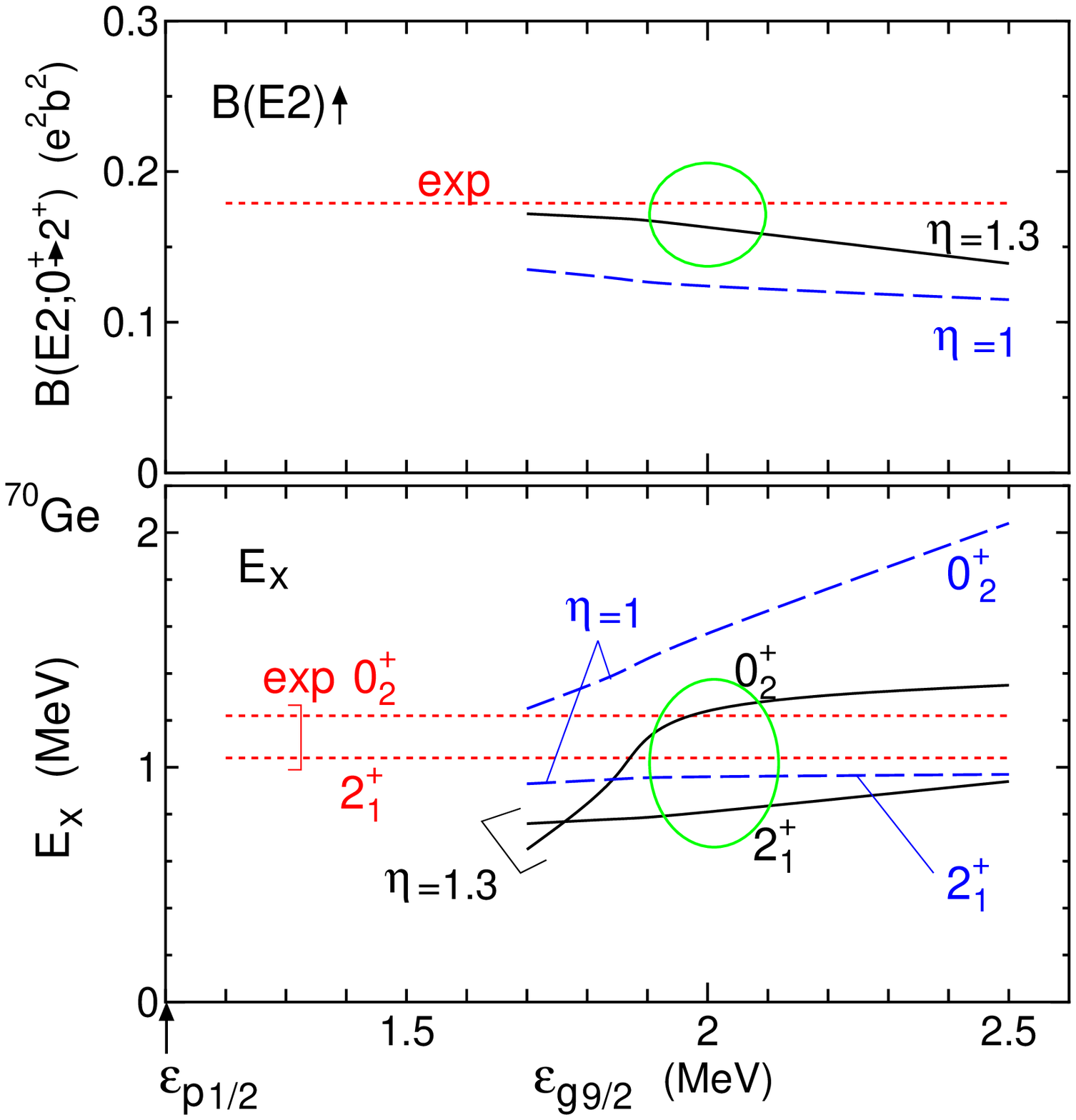}
\includegraphics[width=7.0cm,height=6.4cm]{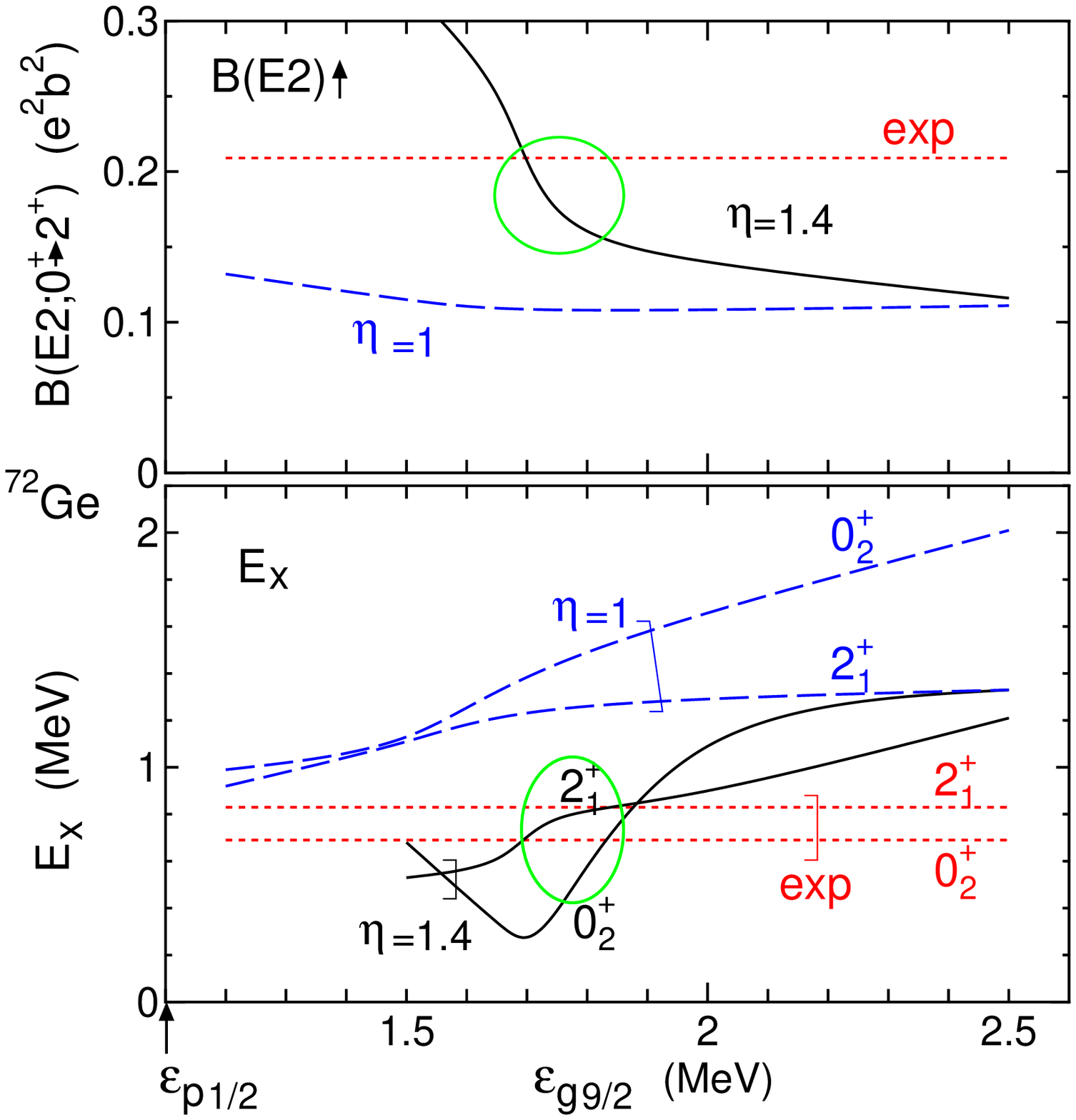}
\includegraphics[width=7.0cm,height=6.4cm]{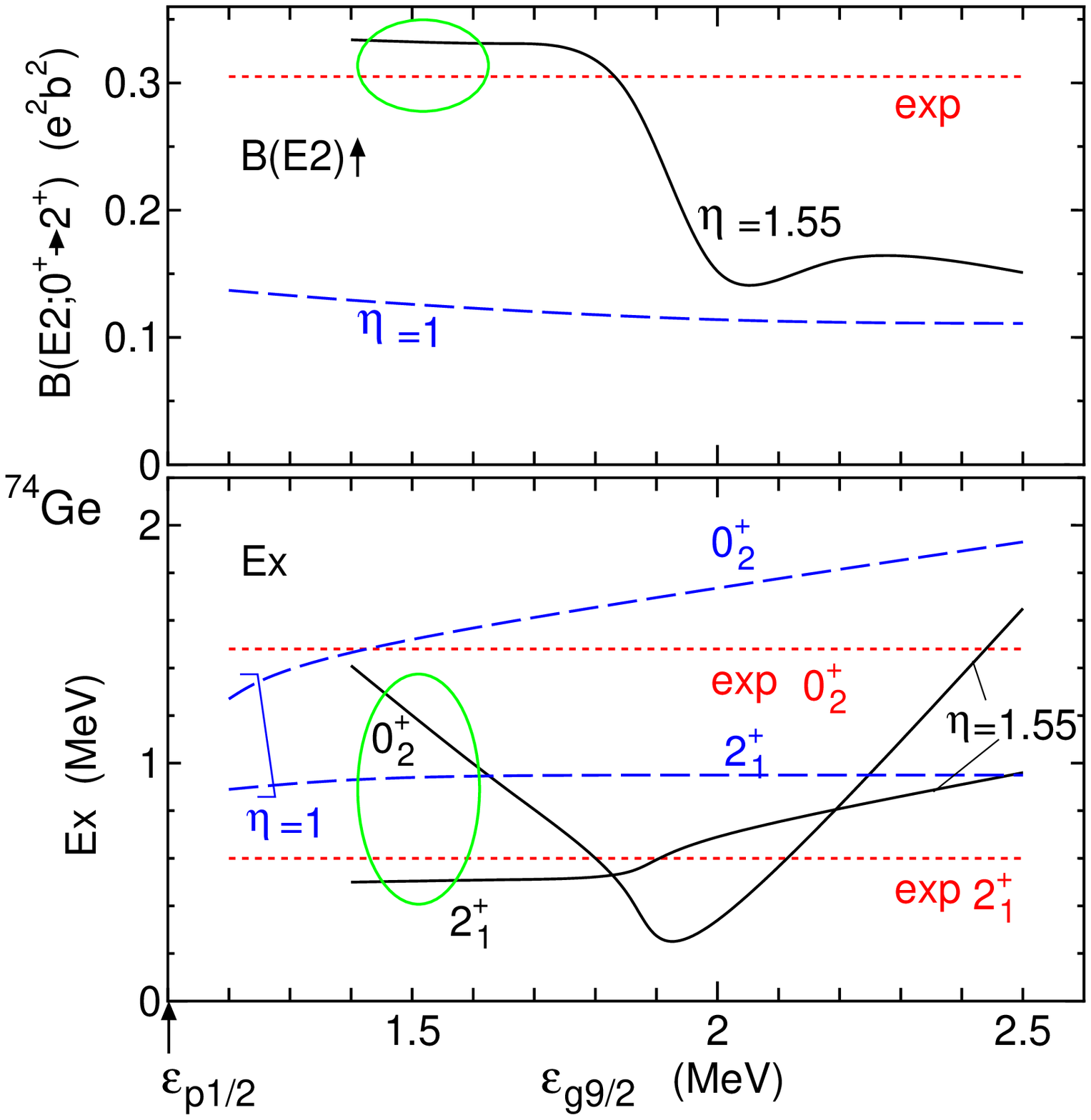}
  \caption{Variations of
           $B(E2)\hspace{-1.5mm}\uparrow$, $E_x(2_1^+)$,
           and $E_x(0_2^+)$ in $^{70,72,74}$Ge as
           $\varepsilon_{g9/2}$ approaches $\varepsilon_{p1/2}$.}
  \label{fig2}
\end{figure}


The effects from the $2d1g_{7/2}3s$ orbits are considered to be
isotope dependent and should become larger as the Fermi level
moves up. Thus in our trial calculations, we assume an empirical
expression for $\eta$, $\eta = 1+ 0.05(N-32)$. With these
considerations, the calculations indeed achieve a great
improvement in excitation energies $E_x(2_1^+)$ and $E_x(0_2^+)$,
especially for $^{66-74}$Ge. The calculated
$B(E2)\hspace{-1.5mm}\uparrow$ values remain good at the beginning
of the chain ($N$ = 34 and 36, because the configurations are
almost exhausted by the $fp$ shell), and are enhanced at the end
with $N$ = 46 $-$ 50. These improvements support the use of
effective quadrupole matrix element $Q_{eff}(g_{9/2}g_{9/2})$.
However, $B(E2)\hspace{-1.5mm}\uparrow$ for $^{72-76}$Ge are again
not enhanced. This indicates that the microscopic content in the
wave-functions has not been improved. As a matter of fact, the
neutron and proton occupation numbers $\langle n_{g9/2}^\nu
\rangle \sim 0.62$ and $\langle n_{g9/2}^\pi \rangle \sim 0.07$
obtained for $^{72}$Ge ($N=40$) by using $Q_{eff}(g_{9/2}g_{9/2})$
cannot bear a comparison with the large values $\langle
n_{g9/2}^\nu \rangle \sim 3$ and $\langle n_{g9/2}^\pi \rangle
\sim 0.2$ reported in Ref. \cite{Langanke2}. The large
$B(E2)\hspace{-1.5mm}\uparrow$ value in $^{72}$Ge corresponds to a
large occupation of the $g_{9/2}$ orbit, which must be due to the
pairing correlations in addition to the $QQ$ force.

In fact, the unusually low energy of the second $0^+$ state in
$^{72}$Ge implies a delicate balance between the pairing
correlations and the energy gap $\varepsilon_{g9/2}-
\varepsilon_{p1/2}$. It has been pointed out by Schiffer {\it et
al.} \cite{Schiffer} and Utsuno {\it et al.} \cite{Utsuno} that
the ``effective energy" of the intruder states changes rapidly
around a closed subshell in neutron-rich nuclei, and these changes
were discussed in terms of the spin-orbit interaction or the
spin-isospin dependence of nucleon-nucleon interaction. In our
case, the change of the effective energy gap $\varepsilon_{g9/2}-
\varepsilon_{p1/2}$ could be attributed to the pairing
correlations in the $1g2d3s$ shell. A good effective Hamiltonian
in a full shell model should be able to describe the balance
microscopically. In the present work with the truncated model
space, we try to understand the problem by simulating the
underlying physics. Namely, we adjust the energy gap
$\varepsilon_{g9/2}- \varepsilon_{p1/2}$ and study any resultant
change in $B(E2)\hspace{-1.5mm}\uparrow$ and $E_x(0_2^+)$.

We study variations of $B(E2)\hspace{-1.5mm}\uparrow$ and
$E_x(0_2^+)$ in $^{66-74}$Ge as a function of single-particle
separation energy between the $g_{9/2}$ and $p_{1/2}$ orbits.
 In the calculation, we use the enhancement factor $\eta =1+
0.05(N-32)$ for $^{66-72}$Ge, and a slightly larger one $\eta
=1.55$ for $^{74}$Ge. Interesting results are obtained as shown in
Fig. \ref{fig2}. The circled areas represent the best choices in
parameter that can describe the observed
$B(E2)\hspace{-1.5mm}\uparrow$, and at the same time, give correct
sequences of $2_1^+$ and $0_2^+$ in $^{70-74}$Ge. Figure
\ref{fig2} indicates clearly the necessity of a rapid lowering of
the single-particle energy $\varepsilon_{g9/2}$ from the original
value 2.5 MeV in $^{66-68}$Ge, to $\sim$ 2.0 MeV in $^{70}$Ge, to
$\sim$ 1.75 MeV in $^{72}$Ge, and to $\sim$ 1.5 MeV in $^{74}$Ge.
Thus the calculation demonstrates that the observed enhancement in
$B(E2)\hspace{-1.5mm}\uparrow$ and variation in $E_x(0_2^+)$
correlate closely with the single-particle energy separation. Data
cannot be understood unless the effective energy gap
$\varepsilon_{g9/2}- \varepsilon_{p1/2}$ near $N=40$ is set to be
small. On the other hand, the calculated
$B(E2)\hspace{-1.5mm}\uparrow$ in $^{72}$Ge and $^{74}$Ge cannot
reach the observed large values unless the enlarged matrix element
$Q_{eff}(g_{9/2}g_{9/2})$ is used. In particular, the correct
sequence of the $0_2^+$ and $2_1^+$ states in $^{74}$Ge cannot be
obtained without narrowing the effective gap $\varepsilon_{g9/2}-
\varepsilon_{p1/2}$ to a sensitive region in the parameter space,
which implies an extremely delicate balance between the pairing
correlations and the energy separation $\varepsilon_{g9/2}-
\varepsilon_{p1/2}$.

\begin{figure}[b]
\begin{center}
\includegraphics[width=7.0cm,height=6.4cm]{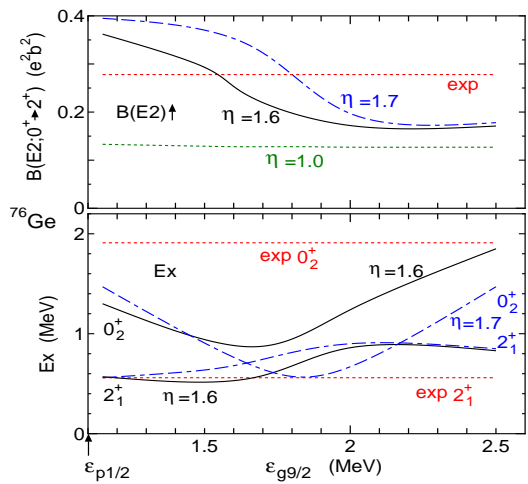}
  \caption{Variations of
           $B(E2)\hspace{-1.5mm}\uparrow$, $E_x(2_1^+)$,
           and $E_x(0_2^+)$ in $^{76}$Ge as
           $\varepsilon_{g9/2}$ approaches $\varepsilon_{p1/2}$.}
  \label{fig3}
\end{center}
\end{figure}

Although a meaningful parameter choice can be found for the
isotopes up to $^{74}$Ge, it is difficult to go beyond them and
find good parameters that can simultaneously reproduce
$B(E2)\hspace{-1.5mm}\uparrow$ and $E_x(0_2^+)$ for $^{76}$Ge, as
illustrated in Fig. \ref{fig3}. This indicates limitations in our
renormalization approach within the truncated model space missing
$2d1g_{7/2}3s$ orbits above $g_{9/2}$. As we shall see in Sect.
IV, a small energy separation
$\varepsilon_{g9/2}-\varepsilon_{p1/2}$ causes nucleons to jump to
the $g_{9/2}$ orbit easily, and with more nucleons in the
$g_{9/2}$ orbit the use of $Q_{eff}(g_{9/2}g_{9/2})$ enhances
$B(E2)$ values in $^{70-74}$Ge. However, the situation seems to be
different between $^{70-74}$Ge and $^{76}$Ge. In $^{70-74}$Ge, the
$g_{9/2}$ orbit is occupied by at most four neutrons. In
$^{76}$Ge, however, about half of the neutron $g_{9/2}$ orbit is
occupied. With an additional neutron pair in the $g_{9/2}$ orbit,
$Q_{eff}(g_{9/2}g_{9/2})$ enhances $B(E2)$ exaggeratedly. In such
a case, our renormalization approach fails to work properly. It
would be necessary for the description of the $N>40$ isotopes to
explicitly include upper orbits such as $d_{5/2}$, as suggested in
Ref. \cite{Caurier}. Below, we continue our discussion but focus
the attention on the structure change in $^{70}$Ge, $^{72}$Ge, and
$^{74}$Ge only.

\begin{figure}[t]
\begin{center}
\includegraphics[width=7.0cm,height=7.0cm]{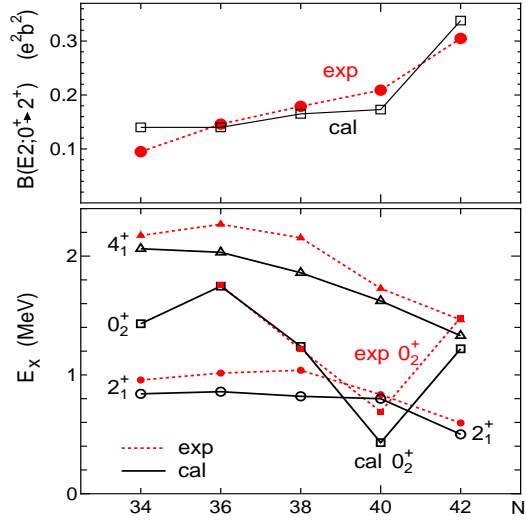}
  \caption{Comparison of improved calculation for
           $B(E2)\hspace{-1.5mm}\uparrow$, $E_x(2_1^+)$,
           and $E_x(0_2^+)$ with experimental data.}
  \label{fig4}
\end{center}
\end{figure}


With the above adjustment in $\varepsilon_{g9/2}-
\varepsilon_{p1/2}$ and the use of effective quadrupole matrix
element $Q_{eff}(g_{9/2}g_{9/2})$, we are able to reproduce
correctly all the observed variations in
$B(E2)\hspace{-1.5mm}\uparrow$, $E_x(0_2^+)$, and $E_x(2_1^+)$ for
$^{66-74}$Ge, as summarized in Fig. \ref{fig4}. Here, the energy
separation $\varepsilon_{g9/2}- \varepsilon_{p1/2}$ is not
required to change for the $^{66}$Ge and $^{68}$Ge calculations if
$Q_{eff}(g_{9/2}g_{9/2})$ with $\eta =1+0.05(N-32)$ is used.
Figure \ref{fig4} shows that the variation trend of $4_1^+$ energy
is also correctly reproduced. Of course, our simulated calculation
does not provide a final answer to the problem, but is quite
suggestive. It should be stressed that the set of circles in Fig.
\ref{fig2} is the sole selection in parameter that can describe
the experimental data.

\section{What changes in structure?}\label{sec4}

\begin{figure}[t]
\includegraphics[width=4.25cm,height=5.0cm]{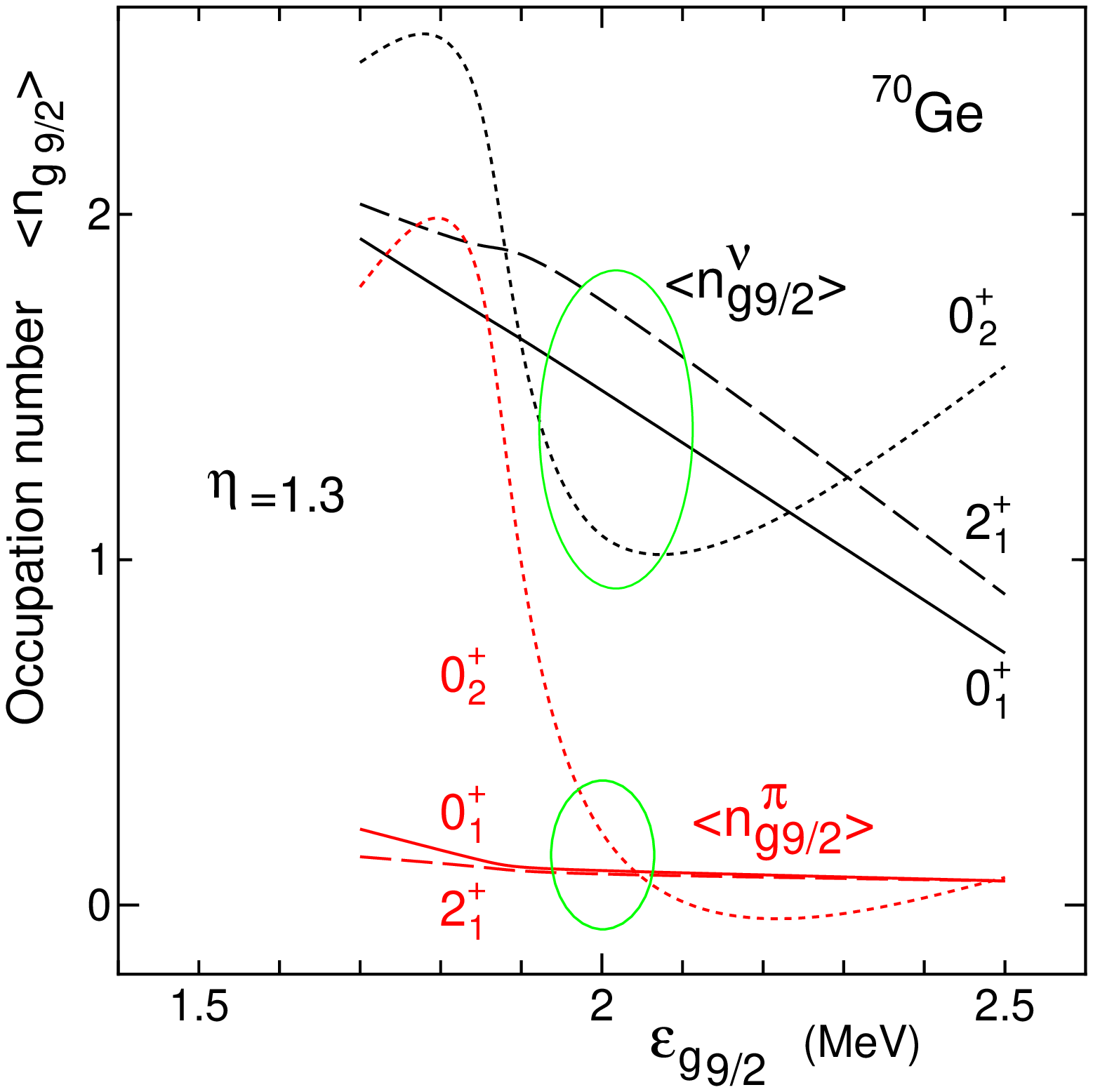}
\includegraphics[width=4.25cm,height=5.0cm]{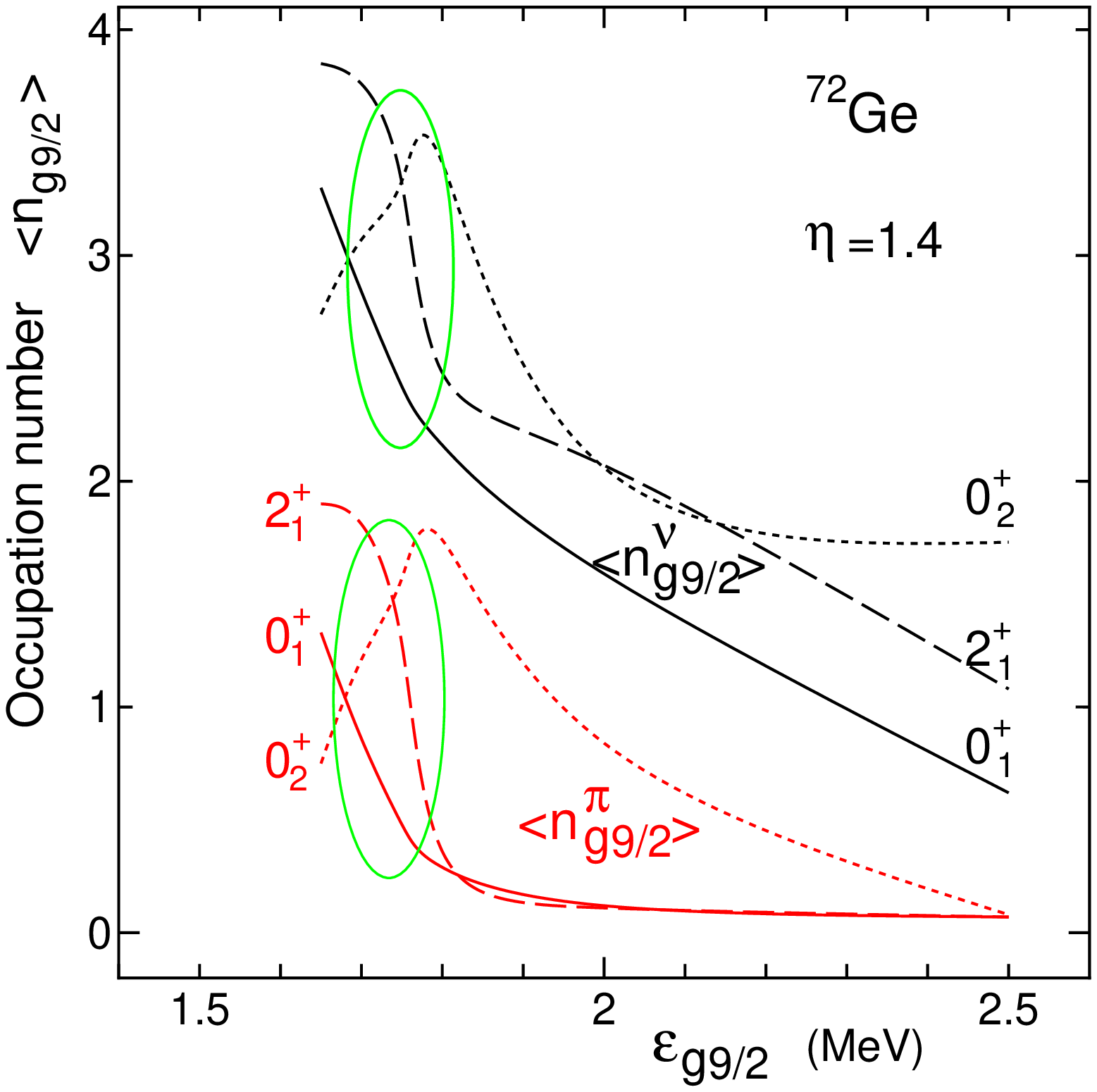}
\includegraphics[width=4.25cm,height=5.0cm]{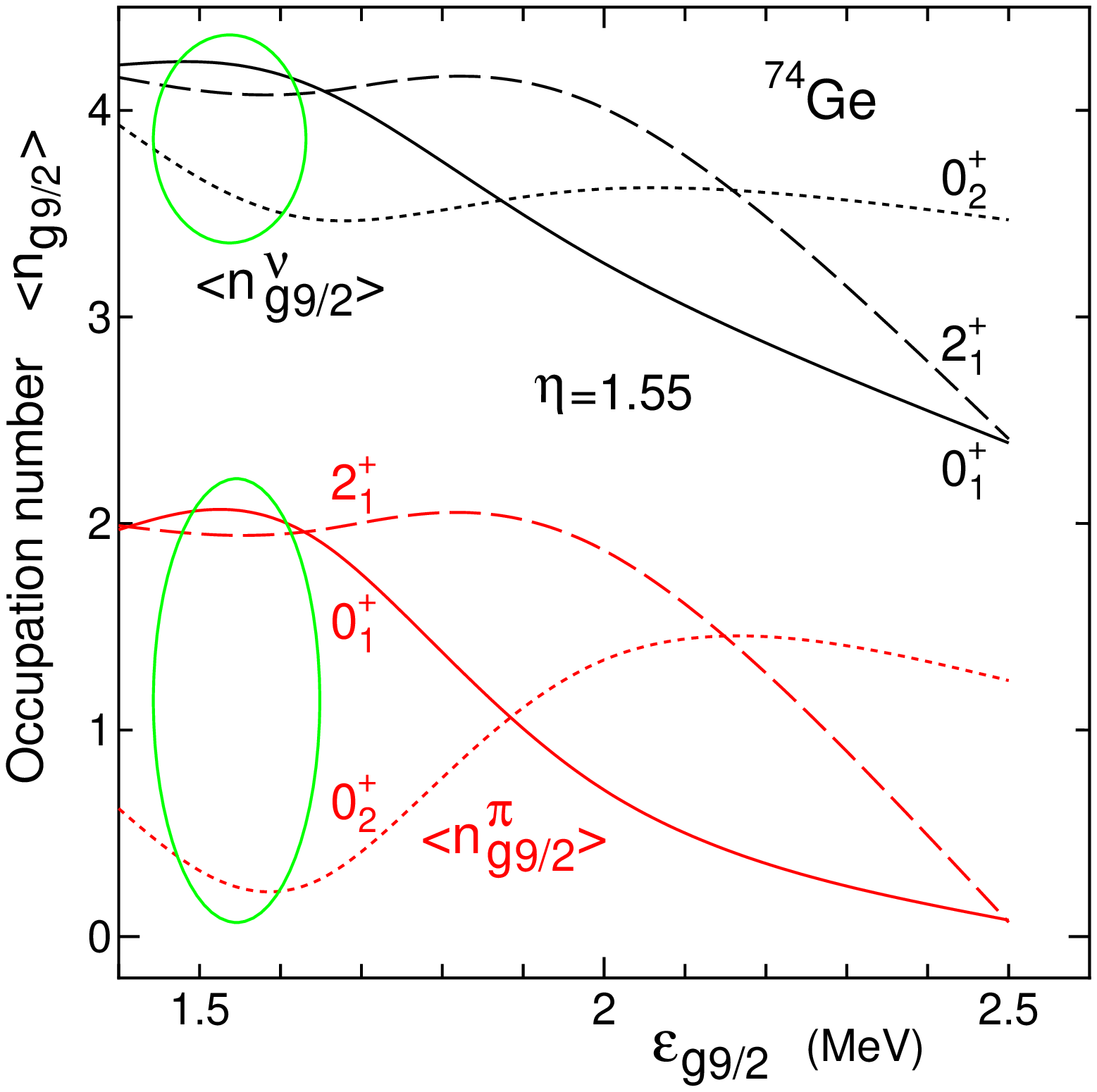}
  \caption{(Color online) Variations of
           $\langle n_{g9/2}^\pi \rangle$ and
           $\langle n_{g9/2}^\nu \rangle$ in $^{70,72,74}$Ge
           as $\varepsilon_{g9/2}$ approaches
           $\varepsilon_{p1/2}$.}
  \label{fig5}
\end{figure}

The important information extracted from the above discussions is
that in order to describe all the observed quantities in Ge
isotopes, it is necessary to have enough $g_{9/2}$ contributions
in the wave-functions. Next, let us study how the $g_{9/2}$ proton
and neutron occupations correlate with the observations and how
they vary in $^{70,72,74}$Ge. We calculate occupation numbers
$\langle n_{g9/2} \rangle$ for protons and neutrons as functions
of the energy separation $\varepsilon_{g9/2}- \varepsilon_{p1/2}$,
and focus our discussion on the results in the circled area that
have reproduced data. Figure \ref{fig5} shows that in $^{70}$Ge, a
considerable number of neutrons occupy the $g_{9/2}$ orbit
($\langle n_{g9/2}^\nu \rangle \gtrsim 1$) while protons are
scarcely found in this orbit. For the ground state of $^{72}$Ge,
the occupation numbers become $\langle n_{g9/2}^\pi \rangle \sim
0.5$ and $\langle n_{g9/2}^\nu \rangle \sim 2.5$. These occupation
numbers are consistent with those of Ref. \cite{Langanke2}, though
our value $\langle n_{g9/2}^\pi \rangle \sim 0.5$ is larger. The
result indicates a superfluid state for neutrons in $^{72}$Ge, in
contrast to the narrow subshell closure in $^{68}$Ni
\cite{Caurier,Sorlin,Kaneko2}. It is notable that in the circled
area for $^{72}$Ge, the coupling between the $0_1^+$ and $0_2^+$
states becomes significant and a considerable number of protons
occupy the $g_{9/2}$ orbit, which shows a qualitative difference
from the structure in $^{70}$Ge. The structure in $^{74}$Ge again
differs qualitatively from $^{72}$Ge. As $\varepsilon_{g9/2}$
approaches $\varepsilon_{p1/2}$, the proton occupation numbers
$\langle n_{g9/2}^\pi \rangle$ for $0_1^+$ and $0_2^+$ are
reversed . For $\varepsilon_{g9/2} \lesssim 1.75$ MeV with which
the calculated $B(E2)\hspace{-1.5mm}\uparrow$ are comparable with
data, more than four neutrons occupy the $g_{9/2}$ orbit and
protons occupy this orbit also considerably, in sharp contrast to
the situation in $^{70}$Ge.

Thus the structure differences appearing in our
calculation for the two lowest $0^+$ states, $0_1^+$ and $0_2^+$,
may offer a new explanation for the old problem observed in
$(p,t)$ and $(t,p)$ reactions. Our simulated calculation predicts
that the proton structure in the $g_{9/2}$ orbit is very different
among the states in $^{70,72,74}$Ge. The proton $g_{9/2}$
occupation changes from a negligible amount in $^{70}$Ge to a
considerable amount in $^{72,74}$Ge, and in addition, with a
different distribution in the $0_1^+$ and $0_2^+$ states.

Obviously, these results cannot be obtained without the presence
of strong proton-neutron correlations in the model, which are
important for nuclei where protons and neutrons occupy the same
shell. The energy difference $\varepsilon_{g9/2}-
\varepsilon_{p1/2} =1.34$ MeV used in our calculation, which is
adopted so as to well describe $^{66-68}$Ge \cite{Hase1} and also
consistent with the value used in Ref. \cite{Aberg}, is rather
small. Still, for the isotopes up to $^{70}$Ge, the proton
occupation number $\langle n_{g9/2}^\pi \rangle$ remains small, in
contrast to the strong configuration mixing within the $fp$ shell
including the $p_{1/2}$ orbit. In this sense, for protons when $N
\le 38$, there exists a gap between the $pf_{5/2}$ shell and
$g_{9/2}$. However, as clearly seen from Fig. \ref{fig5}, the
proton occupation number $\langle n_{g9/2}^\pi \rangle$, together
with the neutron occupation number $\langle n_{g9/2}^\nu \rangle$,
increases suddenly in $^{72}$Ge and $^{74}$Ge. This ``resonance"
effect of $\langle n_{g9/2}^\pi \rangle$ and $\langle n_{g9/2}^\nu
\rangle$ can be best attributed to strong proton-neutron
correlations. Similar effects are obtained in the SMMC
calculations for the $N=40$ isotones \cite{Langanke2}, where a
larger energy difference $\varepsilon_{g9/2}- \varepsilon_{p1/2}
=2.44$ (or 1.97) MeV is used but upper $2d1g_{7/2}3s$ orbits above
$g_{9/2}$ are included in the model space. A considerably large
number of $\langle n_{g9/2}^\pi \rangle$ starts from $^{72}$Ge
(where $Z$ is only equal to 32 and $N$ takes the magic number 40).
We may thus conclude that the proton $g_{9/2}$ occupancy must be
caused by the proton-neutron correlations in the $gds$ shell. Our
treatment with small $\varepsilon_{g9/2}- \varepsilon_{p1/2}$ and
large $Q(g_{9/2}g_{9/2})$ is considered to mimic the effects,
since the mechanism is not properly included in the model. It
should be noted in this regard that our isospin-invariant
Hamiltonian treats dynamically the strong proton-neutron
interactions.

\section{Concluding remarks}\label{sec5}

In conclusion, inspired by the recent experimental advances in
$B(E2)$ measurement for neutron-rich Ge isotopes, we have carried
out a systematical shell model calculation for $^{66-82}$Ge in the
model space $(p_{3/2},f_{5/2},p_{1/2},g_{9/2})$. The calculations
have shown that the strong enhancement of
$B(E2)\hspace{-1.5mm}\uparrow$ and the unusually low excitation of
the second $0^+$ state near $N=40$ can be explained only with
sufficient occupation of protons and neutrons in the $g_{9/2}$
orbit. The simulated calculations that mimic such effects have
suggested a possible understanding of the structure change from
$^{70}$Ge to $^{74}$Ge in terms of rapid increase in the number of
$g_{9/2}$ protons and neutrons. This could be the source for the
long-standing problem that has not yet been understood. The
isotopic dependence of the $g_{9/2}$ proton occupation must have
an origin of strong proton-neutron interactions.

In principle, it is a very difficult question whether all these
effects can be expressed in an effective interaction. The
calculations in Ref. \cite{Caurier} pointed out that the $d_{5/2}$
orbit above the $g_{9/2}$ makes a considerable contribution when
$N>36$ in Cr and Fe isotopes apart from the proton magic number
$Z=28$. The SMMC calculations in the $1f2p-1g2d3s$ space
\cite{Langanke2} also suggested an important role of the $d_{5/2}$
orbit in the enhancement of $g_{9/2}$ occupancies in those nuclei
that have sufficient number of protons in the upper $1f2p$ shell.
According to the extended $P+QQ$ model, we can suggest that the
pairing and quadrupole-quadrupole correlations in the major shell
$1g2d3s$ make it easier for protons and neutrons to be excited
from the $1f2p$ shell to the $g_{9/2}$ orbit. The present
investigation with enlarged $Q(g_{9/2}g_{9/2})$ and decreased
energy gap $\varepsilon_{g9/2}- \varepsilon_{p1/2}$ is along the
line of this thought. The problem may be thoroughly explained if
shell model calculations are performed in a two-major-shell-space
including both $1f2p$ and $1g2d3s$.




\begin{thebibliography} {99}

\bibitem{Padilla} E. Padilla-Rodal {\it et al.}, Phys. Rev. Lett.
 94 (2005) 122501.

\bibitem{Caurier} E. Caurier, F. Nowacki, and A. Poves,
 Eur. Phys. J. A15 (2002) 145 (2002).

\bibitem{Sorlin} O. Sorlin {\it et al.},
 Phys. Rev. Lett. 88 (2002) 092501.

\bibitem{Langanke1} K. Langanke, J. Terasaki, F. Nowacki,
 D. J. Dean, and W. Nazarewicz,
 Phys. Rev. C 67 (2003) 044314.

\bibitem{Langanke2} K. Langanke, D.J. Dean, and W. Nazarewicz,
 Nucl. Phys. A 728 (2003) 109.

\bibitem{Fortune} H.T. Fortune and M. Carchidi,
 Phys. Rev. C 36 (1987) 2584.

\bibitem{Ardouin} D. Ardouin {\it et al.},
 Phys. Rev. C 18 (1978) 1201.

\bibitem{Lebrun} C. Lebrun {\it et al.},
 Phys. Rev. C 19 (1979) 1224.

\bibitem{Mordechai} S. Mordechai {\it et al.},
 Phys. Rev. C 29 (1984) 1699.

\bibitem{Vergnes} M.N. Vergnes {\it et al.},
 Phys. Lett. B 72 (1978) 447.

\bibitem{Lecomte} R. Lecomte {\it et al.},
 Phys. Rev. C 25 (1982) 2812.

\bibitem{Carchidi} M. Carchidi {\it et al.},
 Phys. Rev. C 30 (1984) 1293.

\bibitem{Sugawara} M. Sugawara {\it et al.},
 Eur. Phys. J. A 16 (2003) 409.

\bibitem{Duval} P.D. Duval, D. Goutte, and M. Vergnes,
 Phys. Lett. B 124 (1983) 297.

\bibitem{Hase1} M. Hasegawa, K. Kaneko, and T. Mizusaki,
 Phys. Rev. C 70 (2004) 031301(R);
 71 (2005) 044301; 72 (2005) 064320.

\bibitem{Kaneko1} K. Kaneko, M. Hasegawa, T. Mizusaki,
 Phys. Rev. C 70 (2004) 051301(R).

\bibitem{Hase2} M. Hasegawa, Y. Sun, K. Kaneko, and T. Mizusaki,
 Phys. Lett. B 617 (2005) 150.

\bibitem{Kaneko2} K. Kaneko, M. Hasegawa, T. Mizusaki, and Y. Sun,
 Phys. Rev. C 74 (2006) 024321.

\bibitem{Schiffer} J.P. Schiffer {\it et al.},
 Phys. Rev. Lett. 92 (2004) 162501.

\bibitem{Utsuno} Y. Utsuno, T. Otsuka, T. Glasmacher, T. Mizusaki,
 and M. Honma, Phys. Rev. C 70 (2004) 044307,
 and references cited therein.

\bibitem{Aberg} A. Juodagalvis and S. ${\rm \AA}$berg, Nucl. Phys.
 A 683 (2001) 206.

\end{thebibliography}
\end{document}